\documentclass[aps,preprint,showpacs,epsfig,epsf]{revtex4}
\usepackage{appendix}
\usepackage{graphicx}
\usepackage{amssymb}
\usepackage{amsmath}

\begin{document}
\title { Dynamics of magnetic nanoparticle suspensions} 
\author{Vanchna Singh$^1$, Varsha Banerjee$^1$ and Manish Sharma$^2$}

\affiliation{$^1$Department of Physics, Indian Institute of Technology,
Hauz Khas, New Delhi 110016, INDIA.\\
$^2$Centre for Applied Research in Electronics, Indian Institute of Technology,
Hauz Khas, New Delhi 110016, INDIA.}

\begin{abstract}
We study the dynamics of a suspension of magnetic nanoparticles. Their relaxation times are strongly 
size-dependent. The dominant mode of relaxation is also governed by the size of the particles. As a 
result the dynamics is greatly altered due to polydispersity in the sample. We study the effect of 
polydispersity on the response functions. These exhibit significant changes as the parameters 
characterizing polydispersity are varied. We also provide a procedure to extract the particle size 
distribution in a polydisperse sample using Cole-Cole plots. Further the presence of attractive 
interactions causes aggregation of particles leading to the formation of clusters. Repulsive interactions 
along with thermal disorder not only hinder aggregation, but also introduce the possibility of removal of 
particles or ``fragmentation'' from clusters. The competing 
mechanisms of aggregation and fragmentation yield a distribution of cluster sizes in the steady-state.
We attempt to understand the formation of clusters and their distributions using a 
model incorporating the phenomena of aggregation and fragmentation. Scaling forms for quantities of 
interest have been obtained. Finally we compare our numerical results with experimental 
data. These comparisons are satisfactory.

\end{abstract}
\maketitle
\section{ Introduction} 
\label{sec1}

Single domain magnetic nanoparticles (MNP) and their colloidal suspensions have attracted a lot 
of attention in the recent years \cite{rosen}-\cite{conno}. The growing interest is due to a variety of 
technological applications associated with them.  These range from  mechanical and thermal applications 
involving their usage as sealants, lubricants and coolants to challenging applications in 
medicine for the purpose of  magnetic resonance imaging, targeted drug delivery and biomarkers 
and biosensors to name a few. The main reason behind their wide applicability is the ease 
with which they can be detected and manipulated by the application of an external magnetic 
field. Their response times are strongly size-dependent, thus introducing the possibility of 
synthesizing particles to yield application tailored response times.

Most practical applications require appropriate surfactant coatings to prevent agglomeration
and sedimentation of magnetic nanoparticles. Many  biological and medical applications require 
nanoparticles with biologically relevant coatings in order to use them as probes and carriers 
\cite{conno,astal}. Such coatings enhance the particle size, although the magnetic volume 
remains unchanged. The Neel and Brownian relaxation times which characterize the dynamics of the
suspension depend not only upon the constituting material, but also the magnetic volume and the
enhanced volume due to surfactant coating. The interplay of these parameters governs whether one 
or both relaxation times contribute to the dynamics \cite{rosen}-\cite{coffey},\cite{shlio}. 
In the present paper, 
we systematically analyze the effect of the above parameters on the relaxation times. We also study 
the ac susceptibility $\chi(\omega)$, the most commonly studied response function, in various regimes 
characterized by the relative dominance of the two relaxation times. This understanding makes it 
possible to estimate relaxation times as well as related parameters from the measurement of 
$\chi(\omega)$ in the laboratory.
 
All experimental samples have a distribution of particle sizes and are referred to 
as polydisperse. The particle size distributions in experimental samples, usually 
obtained by a tunnelling electron microscopy (TEM) analysis, are found to have a 
log-normal form \cite{rosen,fannin1}. Hence we incorporate 
the effect of polydispersity in our calculations of $\chi(\omega)$. As the Neel
and Brownian relaxation times have a strong dependence on particle size, polydispersity 
leads to a considerable broadening and sometimes an additional peak in 
$\chi^{\prime\prime} (\omega)$. We have also worked out a procedure to obtain particle size 
distributions from the susceptibility data via Cole-Cole plots when 
they are unavailable \cite{cole1,cole2}. 

The above approach assumes the {\it single particle model} \cite{shlio} applicable to dilute suspensions 
in which magnetic interactions amongst particles may be ignored. Many biological applications are 
particularly benefitted by this approach. For instance diagnostic tools such as magnetic resonance imaging 
rely on transport and manipulation of individual magnetic particles bound by fluorescent dyes through blood 
vessels \cite{zhaom}. However many mechanical and thermal applications rely on viscosity effects arising due 
to a higher density of MNP in the suspensions. When present in sufficient concentration, clustering and 
chaining of MNP is rather common as observed by electron microscopy or dynamic light scattering studies 
\cite{promis}-\cite{kellner}. This behavior is undesirable in many applications using magnetic fluids as 
sealants, coolants, lubricants, printing inks, etc. where invariance of the magnetic and fluid properties 
are paramount \cite{rosen}. On the other hand magnetic domain detection, optical shutters, tagging of
surfaces and other entities benefit from clustering of MNP \cite{rosen,hasmo}. It is hence useful 
to understand the mechanisms responsible for clustering and the dependence of cluster-size distributions
and average cluster sizes on the parameters describing the experimental sample. With this
knowledge, it may be possible to synthesize application tailored suspensions.

Aggregation results due to a variety of interaction energies which come into play in magnetic suspension 
\cite{rosen,odenb,eber1}. Typically these are dipolar, van der Waal and steric interactions. Of these the 
former two are attractive while the later is repulsive. Their relative strengths are governed by the 
magnetic compound, the magnetic volume, the thickness of the surfactant coating as well as the properties 
of the suspending liquid. Additionally, temperature acts as a disordering agent which can not only hinder 
aggregation but also introduce the possibility of detachment of the basic unit(s) from the parent cluster. 
Thus we expect the cluster-size distribution and average cluster size to be governed by the interplay of 
the attractive and repulsive forces amongst the interacting particles and the ambient temperature of 
the suspension. 

Earlier models addressing the issue of clustering in MNP suspensions have treated aggregation as an 
irreversible mechanism. These models obtain a power law growth for the average cluster size in the initial 
stages of aggregation while the steady-state is characterized by a single cluster comprising of all the
particles in the suspension \cite{chandra}-\cite{morim}. The later is often referred to as an infinite
cluster. We further add fragmentation in the model to counter irreversible aggregation, due to the 
presence of repulsive interactions and temperature. The two competing mechanisms make cluster 
formation a self-limiting process resulting in a distribution of clusters of varying sizes with the average 
cluster size governed by the ratio $R$ of the aggregation and fragmentation rates. We obtain the steady-state 
cluster-size distributions and the evolution of the mean cluster size to its steady-state value for 
different values of $R$. Both these functions exhibit scaling. We have compared our results with sets 
of experimental data to lend credence to our interpretations and observations of the aggregation-fragmentation 
model. These comparisons are much to our satisfaction.

The paper is organized as follows. Section~\ref{s2} deals with relaxation mechanisms and response 
functions of dilute MNP suspensions using the single particle model. In Section~\ref{s2_1}, we 
introduce the Neel and Brownian relaxation times and identify regimes where either or both 
relaxation times contribute. Polydispersity and its characterization is introduced in Section~\ref{s2_2}.
The calculation of the ac susceptibility $\chi(\omega)$ for monodisperse and polydisperse samples in the
different regimes described above is presented in Section~\ref{s2_3}. In Section~\ref{s2_4}, we provide a 
procedure to obtain particle-size distributions (for polydisperse samples) using Cole-Cole plots obtained 
from  $\chi(\omega)$. Section~\ref{s3} deals with aspects of clustering occuring in samples which are
no longer governed by the single particle approximation. Interactions prevalent in samples 
exhibiting these phenomena are introduced in Section~\ref{s3_1}. We introduce the 
aggregation-fragmentation model to describe clustering in Section~\ref{s3_2}. The numerical results 
 are presented in Section~\ref{s3_3}. Their comparisons with corresponding measurements in a variety of 
experimental systems are presented in Section~\ref{s3_4}. 
Finally we conclude this paper with a summary and discussion in Section~\ref{s4}. 

\section{Relaxation mechanisms and response functions of dilute MNP suspensions}
\label{s2}
We now study the relaxation properties and the response functions in dilute MNP suspensions
where inter-particle interactions can be ignored. In this regime the particles essentially behave as
independent, single-domain, super paramagnetic entities. 

\subsection{Neel and Brownian relaxation times}
\label{s2_1}
A ferromagnetic sample such as iron for instance, has no net magnetization even below the 
Curie temperature. This is so because the sample comprises of domains, each having  
spontaneous magnetization pointing in a different direction. However if the size of the body 
is reduced, there comes a point beyond which the magnetostatic energy and the energy for forming 
domain walls compete in such a way that a single domain state becomes preferable \cite{sdgupta}.
The direction of magnetization of the single domain particle does not remain
fixed in time though, but undergoes fluctuations or ``relaxations'' as the magnetic moment
rotates between the crystallographic anisotropy axes. As a result the time averaged 
magnetization is still zero and the particle is still paramagnetic. It is called 
``super paramagnetic'' because each particle has a giant magnetic moment arising due to a large
number ($\sim 10^{5}$) of individual atomic moments. We now describe the mechanisms of
relaxation of these single domain super paramagnetic particles suspended in a liquid.

The magnetization vector of a single domain particle is given by \cite{sdgupta}: 
\begin{equation}
\label{spm}
M  = VM_{o}\hat{n},  
\end{equation}
where $V=4\pi r_{c}^{3}/3$ is the magnetic volume of a particle with radius $r_{c}$ usually
reffered to as the core radius, $ M_{o}$ is the saturation magnetization and $ \hat{n}$ 
a unit vector in the direction of the magnetization. In the case of uniaxial anisotropy 
(in the z-direction say), the magnetic energy is given by
\begin{equation}
\label{anis}
E = VK\sin^{2}\theta,
\end{equation}
where $K$ is the effective magnetic anisotropy constant and $ \theta $ is the angle between 
the z-axis and $ \hat{n}$. Minimum energy occurs at $\theta $ = 0 and $\pi$ defining two 
equilibrium orientations corresponding to magnetizations $+VM_{o}$ and $-VM{o}$. If thermal 
fluctuations are strong enough, magnetic moment reversal takes place within the particle by 
overcoming the energy barrier (of height $VK$). This reversal or switching time is called 
Neel relaxation time and is given by
\begin{equation}
\label{neel}
\tau_{N}= \tau_{o} e^{VK/k_{B}T},
\end{equation}
where $\tau_{o}$ is related to the inverse of the attempt frequency of magnetic reversal. 

There is another mechanism by which the magnetic moment of a super paramagnetic particle 
suspended in a fluid can relax. This mechanism of relaxation can be due to the physical 
rotation of the particle within the fluid. It is referred to as Brownian rotational motion as 
it occurs due to the thermal fluctuations in the suspended medium. The Brownian relaxation time is 
given by \cite{coffey}
\begin{equation}
\label{brown}
\tau_{B} = \frac{4 \pi \eta ~{r_{h}}^{3}}{k_{B}T},
\end{equation}
where $\eta $ is the dynamic viscosity of carrier liquid and $r_{h}$ is the  hydrodynamic 
radius defined as the sum of the core radius  $r_{c}$ of the MNP and the surfactant coating 
$\delta$ over it.

As can be seen from Eqs.(\ref{neel}) and (\ref{brown}), both Neel and Brownian relaxation 
times are highly sensitive to the particle size. While $\tau_{N}$ increases exponentially, 
$\tau_{B}$ grows linearly with the particle dimension. It is customary to define an effective
relaxation time as follows \cite{shlio,fannin1}: 
\begin{eqnarray}
\label{teff}
\frac{1}{\tau_{e}} &=& \frac{1}{\tau_{N}} + \frac{1}{\tau_{B}} \quad  \mbox{or} \nonumber\\
\tau_{e} &=&  \frac{\tau_{N}+\tau_{B}}{\tau_{N}\tau_{B}}. 
\end{eqnarray}
Thus it is possible to tailor time scales by an appropriate choice of parameters, particularly $K$,
$r_{c}$ and $r_{h}$. With this in mind, we have systematically studied the effect of $r_{c}$ 
and $r_{h}$ on $\tau_{N}$ and $\tau_{B}$ and consequently $\tau_{e}$ for maghemite 
($Fe_{3}O_{4}$) particles used most commonly in making magnetic fluids. 

In Table I, we summarize our evaluations of $\tau_{N}$, $\tau_{B}$ and $\tau_{e}$ for particles 
with varying magnetic core radius $r_{c}$ and the thickness of the surfactant coating $\delta$. 
In most experiments, the later is usually in the range of 2-6 $nm$. As is observed in Table I, 
$\tau_{B}$ is practically unaffected by $\delta$. For small particles, $\tau_{N}\ll\tau_{B}$ which 
results in $\tau_{e}\approx \tau_{N}$. The relaxation then takes place by rotation of the magnetic 
moment inside the particle. For large particles on the other hand, $\tau_{B}\ll\tau_{N}$. Consequently
$\tau_{e}\approx \tau_{B}$ and the relaxation is due to a physical rotation of the particle 
in the suspension. Thus the choice of the relaxation mode is primarily governed by the particle 
size. For critical particle sizes, often called the cross-over radius $r^{*}$ ($\sim 8 \ nm$ in
Table I), it is found that both mechanisms contribute to the relaxation of the suspended particle. 

We have studied the effect of temperature $T$, the surfactant coating $\delta$ and the anisotropy 
constant $K$ on the cross-over radius $r^{*}$. Most applications require an operating temperature 
in the range of 270 to 320 $K$. We find that $r^{*}$ does not change perceptibly in this range. 
Further, as seen from Table I the variation of $\delta$ does not significatly alter $r^{*}$. The 
anisotropy constant $K$ on the other hand, leads to a substantial change in the corresponding value 
of the cross-over radius. In Figure 1, we plot $r^{*}$ as a function of the anisotropy constant $K$ 
for a few frequently used biological and commercial spherical, magnetic nanoparticles. In all these 
evaluations, the temperature is assumed to be 300 $K$ and the surfactant coating $\delta$ has been 
taken to be 2 $nm$. The figure indicates that the value of the cross-over radius $r^{*}$ is smaller 
for larger values of anisotropy $K$. Thus amongst the three parameters of relevance, the anisotropy 
constant $K$ affects the cross-over radius $r^{*}$ the most.

\subsection{Polydispersity}
\label{s2_2}
Monodisperse samples are an idealization. A variation in the particle size is inherent in all 
experimental samples. TEM studies of several samples have revealed a log-normal distribution for the 
variation of particle sizes \cite{rosen,fannin1,fannin2}. Thus the probability density $P(r_{c})dr_{c}$ 
of having particles within radius $r_{c}$ and $r_{c}+dr_{c}$ can be written as:
\begin{equation}
\label{lognorm}
P(r_{c})dr_{c} =\frac{1}{\sqrt 2\pi\ln\sigma}
	\exp\left[-\ln^{2}(r_{c}/ \bar{r_{c}})/(2 \ln^{2}\sigma)\right]dr_{c},
\end{equation}
where $\bar{r_{c}}$ and $\sigma$ are the mean and variance of the distribution.
Due to the strong dependence of both Neel and Brownian relaxation times on particle size, it is 
evident that a distribution of relaxation times will be obtained if the suspended particles 
have a distribution of sizes. The presence of polydispersity leads to a significant change in the 
behavior of the response function of the sample as we shall see in the following subsection.

\subsection{A C susceptibility  measurements}
\label{s2_3}
The complex susceptibility $\chi(\omega)$ of a suspension of monodisperse MNP in the
linear response regime has the Debye form given by \cite{sdgupta}:
\begin{equation}
\label{debye}
\chi(\omega) = \frac{\chi_{o}}{(1- \iota \omega \tau_{e})}
\end{equation}
where $\chi_{o}$ $=$ $\chi(\omega = 0)$ $=$ $NV^{2}M_{o}^{2}/k_{B}T$ is the static susceptibility of 
the sample comprising of N monodisperse particles of volume V with a saturation magnetization $M_{o}$.
The effective relaxation time $\tau_{e}$ is defined by Eq.~(\ref{teff}).
An experimental time scale is provided by $\omega^{-1}$, the inverse of the frequency of the
applied oscillatory field. If $\omega\tau_{e}$ $\ll$ 1, the response is given by the static
susceptibility $\chi_{o}$ bearing the characteristic $T^{-1}$ dependence. On the other hand,
if the two time scales are comparable, i.e. when $\omega\tau_{e}$ $\approx$ 1, the response is
marked by strong frequency dependent effects. 
Separating the real and the imaginary parts of susceptibility yields:
\begin{eqnarray}
\label{rchi}
\chi^{\prime}(\omega) &=& \chi_{o}~\frac{1}{1+  \omega^{2} \tau_{e}^{2}} \quad \mbox{and}\\
\label{imchi}
\chi^{\prime\prime}(\omega) &=& \chi_{o}~\frac{\omega\tau_{e}}{1+  \omega^{2} \tau_{e}^{2}}.
\end{eqnarray}
The imaginary part of the susceptibility, $\chi^{\prime\prime}(\omega)$ governs dissipation in
the system. When plotted as a function of the frequency, it
exhibits a symmetric peak around $\omega$ = $\tau_{e}^{-1}$. Alternatively, the peak frequency 
can provide information about the radius $r_{c}$ as well as the hydrodynamic radius $r_{h}$ of 
the particle as will be discussed shortly.

The susceptibility response gets substantially altered in the presence of polydispersity.  
For polydisperse samples, $\chi^{\prime\prime}(\omega)$ needs to be 
averaged over the particle size distribution $P(r_{c})$. Thus
\begin{eqnarray}
\label{rchip}
\chi^{\prime}(\omega) &=&\chi_{o}
        \int dr_{c}P(r_{c})\frac{1}{1+\omega^{2}\tau_{e}^{2}(r_{c})} \quad  \mbox{and}\\
\label{imchip}
 \chi^{\prime\prime}(\omega) &=&\chi_{o}
	\int dr_{c}P(r_{c})\frac{\omega\tau_{e}(r_{c})}{1+\omega^{2}\tau_{e}^{2}(r_{c}) }.
\end{eqnarray}
In Figure 2, we plot Eq.~(\ref{imchi}) (open circles) for monodisperse samples corresponding to 
three different values of $r_{c}=$ 4, 8 and 12 $nm$. The chosen values correspond to (a) $r_{c}<r^{*}$, 
(b) $r_{c}\approx r^{*}$ and (c) $r_{c}>r^{*}$ respectively for maghemite particles used to 
generate data of Table I. All cases exhibit the characteristic Debye form. To understand the effect 
of polydispersity, we also plot Eq.~(\ref{imchip}) (filled circles) in the same figure. The 
distribution $P(r_{c})$ vs. $r_{c}$ used for the 
evaluations were obtained from a TEM analysis of maghemite samples used in reference 
\cite{fannin1}. These distributions had a log-normal form with a variance $\sigma \approx$ 0.35. 
The mean value $\bar{r_{c}}$ is the particle size of the corresponding monodisperse evaluation. 

As seen in Figure 2, a broadening of the response function is observed in all cases after the
inclusion of polydispersity. The response no longer has the symmetric  Debye form.
In Figure 2a, the peak frequency provides information about the mean core radius $\bar{r_{c}}$. 
The width of the log-normal distribution often gives rise to a small Brownian 
peak although $\bar{r_{c}}$ $<$ $r^{*}$ and the particles predominantly exhibit Neel relaxation
A pronounced two-peak response is obtained in Figure 2b since $\bar{r_{c}}$ $\approx$ $r^{*}$. 
Both Neel and Brownian relaxation contribute in this regime. The frequencies corresponding to the peak 
values yield information about the average values of core and the hydrodynamic radii
$\bar{r_{c}}$ and  $\bar{r_{h}}$ of the particles. Finally when $\bar{r_{c}}$ $>$ $r^{*}$, the Brownian 
relaxation dominates and the frequency corresponding to the peak provides information regarding the 
hydrodynamic radius $\bar{r_{h}}$ of the particle. Comparing Figures 2a and 2c, it is clear 
that polydispersity affects Neel relaxation more significantly than Brownian relaxation as expected
(cf. Eqs.~(\ref{neel}) and (\ref{brown})).

\subsection{Particle size distributions from Cole-Cole plots}
\label{s2_4}
The information on distribution of relaxation times can be obtained from susceptibility measurements
by empirical models. The most frequently used model for obtaining the distribution of relaxation times
is the Cole-Cole model. Based on this model, the dynamic susceptibility $\chi(\omega)$ with multiple 
relaxation times is given by \cite{cole1,cole2}:
\begin{equation}
\label{chic}
\chi(\omega) = \frac{\chi_{o}}{1 + \left(i\omega\tau_{c}\right)^{1-\alpha}},\\
\end{equation}
where $\tau_{c}$ is the central relaxation time about which all the other relaxation times are
distributed and $\alpha$ is a fitting parameter with limits 0$\le\alpha\le$1. The equation reduces
to the Debye equation for $\alpha$=0. As the deviation from the single relaxation time model becomes
greater, $\alpha\rightarrow$1. Separating the real and imaginary parts of Eq.(\ref{chic}), we obtain:
\begin{eqnarray}
\label{reimc1}
\chi^{\prime}(\omega)&=&\frac{\chi_{o}}{2}
        ~\left(1-\frac{\sinh(1-\alpha)s}{\cosh(1-\alpha)s+\cos(\alpha\pi/2)}\right) \ \ \mbox{and}\\
\label{reimc2}
\chi^{\prime\prime}(\omega)&=& 
                \frac{\chi_{o}}{2}~\frac{\cos(\alpha\pi/2)}{\cosh(1-\alpha)s+\sin(\alpha\pi/2)},
\end{eqnarray}
where $s$ $=$ $\log\omega\tau_{c}$.

Cole and Cole proposed a method of graphically representing the effects of multiple relaxation times.
The method consists of plotting $\chi^{\prime\prime}(\omega)$ for a certain frequency against 
$\chi^{\prime}(\omega)$ at the same frequency. These are called Cole-Cole plots. When $\alpha$ $=$ 0, 
the  Cole-Cole plot is a semi-circle. When $\alpha$ $>$ 0, the Cole-Cole plot is still a semi-circular 
arc similar to a widened Debye curve, but in this case the centre lies below the horizontal axis. The 
plot is symmetrical about the vertical line passing through the point $\chi^{\prime}(\omega)$ $=$ 
$\chi_{o}/2$ when $\chi^{\prime\prime}(\omega)$ is maximum at a frequency $\omega=\tau_{c}^{-1}$. 
The parameter $\alpha$ can be determined from the Cole-Cole plot by a graphical construction. We do not 
reiterate this rather well established procedure here, but refer the reader to reference \cite{cole1} 
for it. It may also be determined by best fits of Eq.~(\ref{reimc1})and Eq.~(\ref{reimc2}) to the 
experimental data. It should be noted that the symmetrical distribution of relaxation times is a 
consequence of a particle size distribution which is symmetric about the mean value $\bar{r_{c}}$. The 
Cole-Cole model assumes a gaussian form for $P(r_{c})$ vs. $r_{c}$. The parameter $\alpha$ is hence 
related to the spread of the gaussian distribution with $\alpha$ $=$ 0 yielding a delta function 
or a monodisperse particle size distribution. 

The response functions in typical MNP suspensions, as observed from Figure 2, are asymmetrical primarily 
due to the log-normal particle size distributions. The Cole- Cole model is thus not the most suitable 
for their description. Experimental systems exhibiting asymmetric response functions  can be 
conveniently represented by an expression due to Cole and Davidson. The ac susceptibility in this model 
is given by \cite{coldavid}:
\begin{equation}
\label{chidc}
\chi(\omega) = \frac{\chi_{c}}{(1 + i\omega\tau_{c})^{\beta}} \ ,
\end{equation}
where $0\le\beta\le 1$ and is related to the deviation from the single relaxation time model. The Debye 
form of Eq.~(\ref{debye}) is recovered when $\beta=1$. Separating the real and imaginary parts of 
Eq.~(\ref{chidc}) results in:
\begin{eqnarray}
\label{rchicd}
\chi^{\prime}(\omega) &=& \chi_{o}\cos\phi^{\beta}\cos\beta\phi \ \ \mbox{and}  \\
\label{imchicd}
\chi^{\prime\prime}(\omega) &=& \chi_{o}\cos\phi^{\beta}\sin\beta\phi,
\end{eqnarray}
where $\phi=\arctan \omega\tau_{c}$. Eqs.~(\ref{rchicd}) and (\ref{imchicd}) give rise to
a ``skewed arc'' in the Cole-Cole plots. The parameter $\beta$ can be obtained by a graphical 
construction or by fitting the experimental data corresponding to $\chi^{\prime}(\omega)$ 
and $\chi^{\prime\prime}(\omega)$ with Eqs.~(\ref{rchicd})-(\ref{imchicd}). 

In Figure 3 we show the Cole-Cole plots for the same polydisperse samples that were used to obtain data 
of Figure 2. The cases (a), (b) and (c) correspond to 
$\bar{r_{c}}<r^{*}$, $\bar{r_{c}}\approx r^{*}$ and $\bar{r_{c}}>r^{*}$ respectively.
In all the three cases, the centre of the arc lies below the horizontal axis due to
multiple relaxation times resulting from polydispersity. Further, the asymmetry in the particle size 
distributions is reflected in the asymmetry of the arcs. The Cole-Cole plots 
are distinct in each of the regimes. In particular, the cross-over regime is characterized by a two-humped 
form signifying a comparable contribution of both  Neel and Brownian relaxation to the response function.  
Thus significant information regarding the experimental sample can be inferred from the Cole-Cole plots. 

We now provide a simple procedure for evaluating the particle-size distribution $P(r_{c})$ vs. $r_{c}$
in the experimental sample when the response function exhibits multiple relaxation times. 
Assuming a log-normal form, $P(r_{c})$ vs. $r_{c}$ is characterized by 
its mean $\bar{r_{c}}$ and variance $\sigma$ signifying the spread in the particle-size 
distributions. While the evaluation of $\bar{r_{c}}$ is straightforward, the variance 
 $\sigma$ needs to be estimated. Recalling
that multiple relaxation times are a consequence of polydispersity, it is imperative to connect
$\beta$ with $\sigma$. In order to find this relationship, we have gone through the following sequence of 
steps. Firstly $\chi^{\prime}(\omega)$ and $\chi^{\prime\prime}(\omega)$ were evaluated using 
Eqs.~(\ref{rchip}) and (\ref{imchip}) for a chosen value of $\bar{r_{c}}$ and $\sigma$.
The corresponding values of $\tau_{c}$ and $\beta$ were then obtained by 
fitting the above susceptibility data with Eqs.~(\ref{rchicd}) and (\ref{imchicd}) of the 
Cole-Davidson model. This procedure was then repeated for a number of 
$\sigma$ values in the range 0.1 to 0.5, keeping $\bar{r_{c}}$ constant. Larger values of $\sigma$ 
were not considered as they resulted in extremely 
broad log-normal distributions which are not of relevance in experiments. 
The above evaluation was then performed for three different values of $K$. 
The results obtained are plotted in Figure 4 on a semi-logarithmic scale. The solid lines follow
the equation: 
\begin{equation}
\label{betasigma}
\sigma = A(K) e^{-B\beta}.
\end{equation}
The parameters $A(K)$ and $B$ in each case are obtained from the best fits of the above equation
to the evaluated data. We find that $B$ $=$ 2.45$\pm$0.05 and is independent of $K$.  
We also find that Eq.~(\ref{betasigma}) and the parameter $B$ is unaffected by $\bar{r_{c}}$, which was varied 
from 4 to 12 nm. 

We test the procedure for extracting particle-size distribution from the susceptibility 
data on maghemite nanoparticle suspensions provided in reference \cite{fannin1}. From the 
Cole-Cole plots of this data, we 
evaluated the best-fit values of $\tau_{c}$ and $\beta$. These were found to be $2\times 10^{-8}$ s and
0.44 respectively. The average core radius $\bar{r_{c}}$, evaluated
from $\tau_{c}$ using Eq.~(\ref{neel}), was found to be 4.2 $nm$. 
The value of $\sigma$ characterizing the spread of the log-normal distribution was then obtained 
using Eq.~(\ref{betasigma}) and was found to be 0.36. The reconstructed particle-size distribution
is shown in Figure 5. The particle-size distribution obtained using TEM measurements has also been 
provided in the figure for comparisons. 
In general, we find that this procedure reproduces the original distributions rather well when the 
response function has one dominant peak, be it in the Neel or the Brownian relaxation regime. 

\section{Clustering in MNP suspensions}
\label{s3}
The properties of the magnetic fluid are greatly affected by the aggregation of particles
(inspite of surfactant coating) due to the presence of attractive and repulsive
interactions. It is hence instructive to consider the different interaction energies in MNP 
suspensions and understand their interplay which leads to the formation of aggregates or 
clusters of different sizes. 

\subsection{Interactions in MNP suspensions}
\label{s3_1}
The primary interaction energies in these systems are enumerated below \cite{rosen}:
\begin{enumerate}
\item Dipolar interaction: \\
As each particle is magnetized, the attractive dipole-dipole interaction could in principle
force particle agglomeration. The dipolar interaction energy between two magnetic
nanoparticles, each having a magnetic moment $\mu$ is given by:
\begin{equation}
\label{dipole}
E_{d}(s)= -\frac{\mu_{o}}{4~\pi}\left(\frac{3~\mu\cdot~(\mu\cdot \vec{s})\vec{s}}{s^5} 
-\frac{\mu ^2}{s^3}\right),
\end{equation}
where $s$ is the center-to-center separation between the two nanoparticles and the permeability 
of free space $\mu_{o}$ $=$ $4\pi\times10^{-7}$ $H m^{-1}$. As the magnetic properties
of the particles are affected by temperature, the dipolar interaction defined above is
also temperature dependent. 

\item van der Waal's interaction:\\
It arises spontaneously between neutral particles because of the fluctuating electric 
dipolar forces and is attractive in nature. Hamaker calculated this interaction for identical 
spheres separated by a surface-to-surface distance $l$ to be:
\begin{equation}
E_{v}=\frac{A}{6}\left\{\frac{2}{l^2+4~l}+\frac{2}{(l+2)^2}
		+\ln\left(\frac{l^2+4l}{l^2+4l+4}\right)\right\}.
\end{equation}
In the above equation, the Hamaker constant $A= 10^{-19} \ Nm$. It should be noted that the $l^{-1}$
dependence in the above equation indicates that infinite energy is required to separate a
particle pair while a finite energy is required in its formation. Therefore agglomeration
of particles will occur as long as the Hamaker constant has a finite value. As a result 
van der Waal's interaction unlike dipolar interaction is unaffected by temperature.

\item Steric interaction:\\
The steric energy comes into play due to the presence of long chained surfactant molecules
coating the particles. This mechanism prevents the particles from approaching very close to
one another, thus preventing the van der Waal's attraction to come into play. 
This repulsive interaction originates due to a compression of the
surface-adsorbed surfactant molecules when the inter particle separation is smaller than
two times the thickness of the surfactant layer. For a sufficient density of the surfactant
layer, the repulsion can grow large enough to avoid the contact between the magnetic particles.
This repulsive energy for spherical particles has been calculated by Papell and is
given by the following form:
\begin{equation}
\label{steric}
E_{s}=\frac{E}{1.325}\left(2\delta-\frac{s}{2}\right)^{\frac{5}{2}}
		~\left(\frac{d_{c}}{2}+\delta\right)^{\frac{1}{2}}.
\end{equation}
In the above form, the Young's modulus $E$ was assumed to be $10^{6}$ and $d_{c}=2r_{c}$ is
the diameter of the magnetic core of the particle.
\end{enumerate}.

Apart from the above interaction energies, the thermal energy responsible for the Brownian motion
of the suspended particles also plays an important role in the aggregation dynamics. An appropriate 
value of temperature (which depends on the size of the interacting particles) is capable of preventing 
aggregation. In principle, this condition may be expressed as a comparison between the attractive 
energy with the thermal energy of the two particles which yields a numerical value of the core 
radius $r_{c}$ at which this could happen. We expect the thermal degrees of freedom, in conjunction with 
steric repulsion, to not only hamper the aggregation process but also remove particles from the parent 
cluster leading to its fragmentation. 

At this juncture we define a parameter $\Omega$ as the ratio of energies leading to aggregation to
those which lead to fragmentation follows:
\begin{equation}
\label{eta}
\Omega = \frac{E_{d}+E_{v} }{E_{s}+k_{B}T}.
\end{equation}
It is expected that the distribution of cluster sizes and the mean cluster size will be governed 
by the ratio $\Omega$. In the following subsection, we present a model which incorporates 
the competing mechanisms of aggregation and fragmentation.  

\noindent\subsection{The aggregation-fragmentation model}
\label{s3_2}
The formulation introduced by Smoluchowski is especially useful to model a suspension of magnetic 
nanoparticles \cite{chandra}. To begin with, we assume that the suspension contains $N$ identical, single
particles executing Brownian motion. The later leads to aggregation of two particles if they come 
within an appropriate range of one another and the net interaction between them is attractive. The 
resulting cluster of size two also executes Brownian motion, but with a reduced diffusion rate till 
it encounters a particle or a cluster of particles. The process goes on and eventually a single large 
cluster comprising of all the $N$ particles is formed. Such a cluster is usually reffered to as an 
infinite aggregate. In most useful suspensions however there is a distribution of clusters of varying 
sizes. As discussed in the preceding subsection, the combined effects of thermal energy and 
inter-particle repulsion can introduce fragmentation in the cluster dynamics
thereby preventing the formation of large aggregates. We thus include this additional mechanism 
in the rate equations which describe the evolution of clusters. 

Let $c(k,t)$ denote the number of clusters containing $k$ particles at time t. The time evolution
of $c(k,t)$ is governed by the following rate equations:
\begin{eqnarray}
\label{reqn}
\frac{\partial c(k,t)}{\partial t} &=& 
	\frac{1}{2}\sum_{i+j=k}K_{ij}c(i,t)c(j,t) - c(k,t)\sum^{\infty}_{j=1} K_{kj}c(j,t)\nonumber\\
	& & + f_{k+1}c(k+1,t)- f_{k}c(k,t) +\delta_{k,1}\sum^{\infty}_{j=1} f_{j}c(j,t), \quad  k \ge 1.
 \end{eqnarray}
In the above equation the $K_{ij}$ and $f_{k}$ are the aggregation and fragmentation kernels respectively
whose forms are specified below.   The aggregation kernel describes the coalescence of a cluster containing
$i$ particles with another containing $j$ particles to yield a larger aggregate comprising of $k$ $=$ $i+j$
particles. It is assumed to have a mass-dependent form defined by $K_{ij}$ $=$ $D(i^{-\mu}+j^{-\mu})$ to 
take into account the reduced mobility of large clusters. A value of $\mu$ $=$ 0 implies a mass-independent 
mobility, i.e., clusters of all sizes diffuse with the same ease. 
A non-zero values of $\mu$ results in a slower mobility of larger clusters and consequently a slower growth
rate of clusters. In the limit of $\mu=\infty$, only monomers are mobile. The choice of $\mu$ is dictated by the 
experimental parameters as we shall see in a short while. The fragmentation kernel $f_{k}$ describes
the loss of a particle from the parent cluster and can also be assumed to have a mass-dependent form defined 
by $f_{k}$= $\omega~k^\nu $. In our studies however, we have set $\nu $ $=$ 0 to reduce the number of 
parameters in the model. The parameters $D$ and $\omega$ define the relative strength of the
aggregation and fragmentation processes. 

The first and the third term in Eq.(\ref{reqn}) are referred to as the gain terms which result in 
the formation of clusters of size $k$. The former describes aggregation of two clusters to yield a 
cluster containing $k$ particles while the later describes the generation of a cluster comprising of $k$ 
particles due to fragmentation of a particle from a cluster of size $k+1$. The second and the fourth 
term on the other hand are referred to as the loss terms which describe processes leading to loss of 
clusters of size $k$. This could be due to aggregation of a cluster of size $k$ with another or the 
its fragmentation. The fifth term in the equation describes the generation of single particles from 
clusters due to the process of fragmentation.  It is easy to check that the following sum rule is 
satisfied:
\begin{equation}
\label{srule1}
\frac{\partial}{\partial t}\left(\sum_{k=1}^{\infty}kc(k,t)\right)= 0, \ \ \mbox{or} \ \ 
\sum_{k=1}^{\infty}kc(k,t) = N,
\end{equation}
as required by conservation conditions.

The condition $\partial c(k,t)/\partial t=0$ describes the steady-state which is of interest to us. In 
the absence of the fragmentation term, Eq.(\ref{reqn}) reduces to the Smoluchowski equation describing 
coagulation phenomena \cite{chandra}. There have been a few studies of this model, both analytical and 
numerical, to predict scaling forms associated with cluster growth and cluster-size distributions
\cite{meakin}-\cite{morim},\cite{ernst}-\cite{kalachev}. The growth of clusters with time 
is a power governed by the relation $<k(t)>$ $\sim$ $t^{z}$. Choosing $\mu$ $=$ 1, the value of $z$ was 
estimated to be $\simeq$ 0.5. The steady state in this model was found to be an infinite (single) aggregate 
comprising of all the $N$ particles. Note that in the absence of fragmentation, the constant $D$ can be absorbed 
by redefining $t$ as $Dt$ in Eq.~(\ref{reqn}), making the later independent of the aggregation rate.  

To mimic those physical situations which do not 
have an infinite-aggregate as a steady-state but rather have a distribution of clusters of varying sizes, 
it is essential to include fragmentation as a competiting mechanism to aggregation. Of relevance in 
the context of the present study is model incorporating  mass-independent aggregation and mass-dependent 
evaporation processes reported in reference \cite{krapi}. The competition between aggregation and evaporation 
leads to several asymptotic outcomes of the steady-state solution of this model. For instance if evaporation 
dominates over aggregation, the steady-state cluster size distribution $c(k)$ vs. $k$ decays exponentially. 
On the other hand for a critical evaporation rate, the distribution decays as $k^{-5/2}$. 

It is convenient to rewrite Eq.(\ref{reqn}) in terms of $P(k,t)$, the probability of having cluster 
of size $k$ at time $t$. Defining the later as $P(k,t)=c(k,t)/\sum_{k=1}^{\infty}c(k,t)$, 
its evolution is governed by the following equation:
\begin{eqnarray}
\label{req2}
\frac{\partial P(k,t)}{\partial t} &=& \sum_{i+j=k} K_{ij}~P(i,t)P(j,t) 
	-P(k,t)\sum^{\infty}_{j=1}K_{kj}P(j,t)\nonumber\\
	& & +\omega P(k+1,t)-\omega P(k,t), \ \ k > 1,\\
\label{req21}
\frac{\partial P(1,t)}{\partial t} &=& - P(1,t)\sum^{\infty}_{j=1} K_{kj}P(j,t)
			+\omega\sum^{\infty}_{j=2} P(j,t), \ \ k = 1.
\end{eqnarray}
The above equations also satisfy the sum rule 
\begin{equation}
\label{sumrule}
\frac{\partial}{\partial t}\left(\sum_{k=1}^{\infty}P(k,t)\right) = 0, \quad \mbox{or} \quad \sum_{k=1}^{\infty}P(k,t) = 1, 
\end{equation}
as required. 

\subsection{Numerical results}
\label{s3_3}
We now solve the the set of equations defined by Eqs.~(\ref{req2})-(\ref{req21}) numerically to obtain the 
steady-state cluster size distribution $P(k)$ vs. $k$. It is useful to define the ratio $R$ characterizing the
relative strengths the aggregation and fragmentation mechanisms:
\begin{equation}
\label{ratio}
R = \frac{D}{\omega}.
\end{equation} 
Identifying the physical origin of $D$ in the attractive interactions between clusters and that of 
$\omega$ in the disordering agents (such as repulsive interactions and temperature), we can expect
$R$ to have the same qualitative effect  as the ratio $\Omega$ defined in Eq.~(\ref{eta})  on steady-state 
cluster size distributions. We use this correspondence to bring contact between numerical 
results and experimental observations on cluster formation in MNP suspensions.

In Figure 6a, we look at the variation of the mean cluster size $<k(R,t)>$ as a function of $t$ for
different values of $R$ on a double logarithmic scale. The parameter $\mu$ was chosen 
to be 2.0. After an initial growth period obeying  a power law, the cluster size attains a steady-state 
value of $<k(R)>$ due to the competing mechanisms of aggregation and fragmentation. As expected, 
$<k(R)>$ increases with  increasing values of $R$. In fact we find that $<k(R)>$ $\sim$ $R^{\alpha}$ with 
$\alpha$ $=$  0.85 $\pm$0.02 as can be seen in the inset. Further, in the diffusion domiated regime 
($R \gtrsim 0.5$) the  data 
in Figure 6a can be scaled by replotting  $<k(R,t)>/<k(R)>$ vs. $t/t_s$ where $t_{s}$ is the time taken 
to attain the steady-state. The initial cluster growth is of the form $<k(t)>\sim t^{z}$. We find that 
the growth exponent $z$ $=$ 0.38 $\pm$ 0.02 . The scaled data is shown in Figure 6b on a double logarithmic 
scale.  The best fit line with a slope of 0.38 is also indicated. The value of the growth exponent also 
depends upon the value of $\mu$. For $\mu$ $=$ 1.0, our simulations yield $z$ $=$ 0.78 $\pm$ 0.02. 
The corresponding scaled data and the power law fit is also shown in the inset of Figure 6b. The faster 
growth of clusters is a consequence of increased mobility due to a lower value of $\mu$.

Next in Figure 7a we plot the steady-state distribution $P(k)$ vs. $k$ corresponding to $\mu$ $=$ 2 for 
the same set of $R$ values considered above. We find that the tails of the distributions fit 
well to a power law in $k$. In Figure 7b, we plot $<k(R)>P(k)$ vs. $k/<k(R)>$ where $<k(R)>$ is the steady-state 
average cluster size for the corresponding value of $R$. These plots indicate that the distribution 
functions $P(k)$ corresponding to different values of $R$ obey scaling in the diffusion-dominated 
regime. The scaling relation can be summarized in the following equation:
\begin{equation}
\label{scaling}
P(k) = \frac{1}{<k(R)>}f\left(\frac{k}{<k(R)>}\right), \quad <k(R)>\sim R^{\alpha}.
\end{equation}

\subsection{Comparisons with experimental data}
\label{s3_4}

We now compare our simulation results with experimental data on average cluster sizes in a variety 
of magnetic nanoparticle suspensions. Eberbeck, et al studied the aggregation of various magnetic 
nanoparticles in a variety of media such as water, phosphate buffered saline, calf serum, bovine 
serum and human serum \cite{eber1,eber2}. The experimental measurements were performed at 
room temperature. In all cases, formation of dimers and trimers were reported by the authors. To make 
comparisons with the simulation results of our model, we first calculate the ratio $\Omega$ defined 
in Eq.(\ref{eta}) for each of the samples. These along with the sample specifications are 
tabulated in Table II. As can be seen in column 4, $\Omega$ is in the range 0.35 to 0.70. Referring to 
the plot of $<k(R)>$ vs. $R$ for $\mu$ $=$ 2 in the inset of Figure 6a, a value of $R$ in the above range yields 
dimers and trimers in the steady-state. 

Next we look at experiments which report the formation of long chains of magnetic
nanoparticles in the presence of an applied magnetic field $H$. The particles acquire dipole moment 
$\tilde{\mu}$ given by: 
\begin{equation}
\label{mueff}
\tilde{\mu} = \frac{4}{3}\pi r_{c}^{3}\mu_{o}\chi H,
\end{equation}
where $r_{c}$ is the core radius of the magnetic particle and $\chi$ is the magnetic susceptibility. 
The interaction energy between two particles with aligned, identical dipole moments is given by:
\begin{equation}
\label{dipolen}
U(s,\theta) = \frac{\mu^{2}}{4\pi \mu_{o}} \frac{1-3cos^{2}\theta}{s^{3}},
\end{equation} 
where $s$ is the centre to centre separation between the two particles and $\theta$ is the angle between 
the applied field and the line joining the centres of the spherical particles. The interaction is thus 
attractive when the dipoles are head-to-tail and repulsive when they are side-by-side. For these samples,
the dominant interaction is dipolar. Thus the parameter $\Omega$ defined in Eq.~(\ref{eta}) reduces to
\begin{equation}
\label{etasf}
\Omega_{H} = \frac{\pi r_{c}^{3}\mu_{o}\chi^{2} H^{2}}{9k_{B}T},
\end{equation}
where we have assumed the dipolar interaction to be at its maximum, when the particles are aligned with
the external field. In the presence of an external field, if the resulting magnetic
dipole-dipole interaction between particles exceeds the thermal energy, then the particles arrange
themselves into chains which are column-like for low volume fractions and worm-like structures for
high-volume fraction \cite{promis,martin}. Many theoretical studies on the above systems have been 
based on Smoluchowski kinetic equations for irreversible aggregation \cite{meakin,kandel}. 


A set of experimental measurements which we find especially relevant in the context of the 
aggregation-fragmentation model are reported in reference \cite{promis}. In this paper, the authors have 
studied aggregation dynamics in very dilute emulsions of ferrofluid droplets in water. The ferrofluid 
droplets were small $Fe_{3}O_{4}$ grains in kerosene coated with a surfactant to prevent 
agglomeration. The data on average chain lengths as a function
of time for different volume fractions and applied fields was obtained using dynamic light scattering 
experiments. We reproduce a scaled form of this  data in Figure 8. On the x-axis, we plot $t/t_{s}$ 
where $t_{s}$ is the time taken to reach the steady-state value of the average cluster. The y-axis has 
been scaled by the steady-state value of the average chain length. The unscaled data is also provided
in the inset for reference. As can be observed in Figure 8, the scaled form (as well as the unscaled form) of the 
experimental data bears a qualitative resemblance to Figure 6b which results from the numerical solution 
of the aggregation-fragmentation model defined by Eqs. (\ref{req2}) and (\ref{req21}). The slow initial 
growth, the power law form at intermediate times and saturation to a steady-state value as time progresses 
are bourne by both sets of data. 

On the quantitative side, we find that the experimental data is well represented by a growth exponent 
$z$ $=$ 0.78$\pm$0.02 obtained with $\mu$ $=$ 1 in our simulations. The corresponding line is
depicted in figure 8.  It is pertinent to 
recall here that the dipolar interaction becomes dominant in the presence of the applied magnetic field 
thereby increasing the diffusivity of the clusters. In the aggregation-fragmentation model, the later can 
be achieved by an increase in the value $D$ as well as a decrease in the value of $\mu$. Consequently, 
we find that our data on $<k(R,t)>$ vs. $t$ for a value of $\mu$ $=$ 1 rather than $\mu$ $=$ 2 results in 
accurate comparisons with experimental data. The initial growth does not conform to the predicted value
of $z$ $=$ 0.5 (also obtained with $\mu$ $=$ 1) of the irreversible aggregation model.

A similar set of experimental results were obtained in reference \cite{martin} where the power law 
prediction for cluster growth was checked by performing a large number of experiments on aqueous 
solutions of super paramagnetic polystyrene beads having a uniform distribution of $Fe_{3}O_{4}$ 
particles. The data was obtained for five values of the volume fraction in the presence of low 
field strengths and higher field strength using optical microscopy. The exponent $z$ describing the 
cluster growth in these studies was found to be much larger than the predicted value of 0.5 especially 
for small volume fractions and low field strengths. These observations further reiterate the 
appropriateness of introducing fragmentation along with aggregation especially in the above limits 
where the thermal energy plays a significant role in the dynamics.

\section{Summary and Discussion}
\label{s4}
We conclude this paper with a summary of results and discussion presented here. Our main interest
was to understand the factors governing the dynamical response of suspensions of single-domain
magnetic nanoparticles. Such an understanding introduces the possibility of synthesizing particles
with application tailored response times. The effect of sample parameters on the Neel and 
Brownian relaxation times which characterize the response was studied. Amongst all the parameters 
of relevance,  the anisotropy of the constituting material and the particle size alter the 
relaxation time most significantly. Infact the dominant relaxation time is also decided primarily
by the particle size. We also studied how these parameters affect the ac susceptibility $\chi(\omega)$ 
which is the most commonly studied response function in the laboratory. This understanding 
proves to be useful in estimating relaxation times as well as sample parameters from the measurement
of $\chi(\omega)$ in the laboratory. 

We have then studied the effect of polydispersity, an inherent feature of all samples on the
response characteristics. These exhibit significant changes due to the strong dependence of
relaxation times on particle sizes. The primary effect is the broadening of response functions
and in some cases the later exhibits a two peaked structure. We have also worked out a procedure to 
obtain the particle-size distribution from  $\chi(\omega)$ using Cole-Cole plots and 
the analysis of Cole and Davidson. This provides an alternative approach to TEM analysis 
which is usually employed to obtain particle-size distributions.

The above studies assumed a single-particle model applicable to dilute suspensions. 
However in many cases the inter-particle interactions cannot be ignored. Clustering
of particles is inevitable in such suspensions. While the formation of clusters is 
undesirable in some applications, it is beneficial in many others. Hence we have tried to 
understand the mechanisms responsible for clustering and the experimental parameters
which govern the properties of cluster-size distribution and the average cluster size.
This knowledge is useful for the synthesis of application tailored suspensions.

A model incorporating the phenomena of aggregation and fragmentation was used to understand 
aspects of clustering. The steady-state cluster size distributions of the model were obtained
by numerically solving the rate equations describing the evolution of clusters. We have 
obtained scaling forms for the cluster-size distributions and the average cluster size. 
Our results agree well with experiments where clustering or chain formation have been observed. 

\noindent {\bf Acknowledgment} VS and VB would like to acknowledge the support 
of CSIR Grant No. 03(1077)/06/EMR-II. VB would also like to acknowledge the hospitality of
ICTP, Italy where part of the work was completed.

\newpage

\begin{center}
{\bf Figure Captions}
\end{center}
\noindent{\bf Table I:} Variation of the Neel ($\tau_{N}$), Brownian ($\tau_{B}$) and effective 
($\tau_{e}$) relaxation times as a function of the core radius $r_{c}$ of the magnetic nanoparticle. 
$\tau_{B}$ has also been evaluated for three values of the surfactant coating $\delta$. The 
corresponding $\tau_{e}$ for these values is also evaluated.\\

\noindent{\bf Table II:} Calculated value of the ratio $\eta$ and the corresponding estimated value
of the cluster size for a variety of MNP used in experiments by Eberbeck, et al \cite{eber1,eber2}. The 
experimental parameters associated with each sample are also specified.\\  

\noindent{\bf Figure 1:} Variation of the cross-over radius $r^{*}$ as a function of the anisotropy 
constant $K$ for several biologically and commercially used magnetic nanoparticles. The surfactant
coating $\delta$ and the temperature $T$ were taken to be 2 $nm$ and 300 $K$ respectively in these
evaluations.\\

\noindent{\bf Figure 2:}  Variation of $\chi^{\prime\prime}(\omega)$ vs. $\omega$ for monodisperse 
(open circles)and polydisperse (filled circles) manghemite samples when (a) $\bar{r_{c}}<r^{*}$ and Neel
relaxation dominates, (b) $\bar{r_{c}}\approx r^{*}$ and
both Neel and Brownian relaxation contribute and (c) $\bar{r_{c}}>r^{*}$ and Brownian relaxation 
dominates. The values of $\bar{r_{c}}$ in (a), (b) and (c) have been chosen as 4, 8 and 12 $nm$ 
respectively.  The surfactant
coating $\delta$ and the temperature $T$ were taken to be 2 $nm$ and 300 $K$ respectively in these
evaluations.\\

\noindent{\bf Figure 3:} Cole-Cole plots (open circles) for data of figure 2 when (a) $\bar{r_{c}}<r^{*}$ 
and Neel relaxation dominates, (b) $\bar{r_{c}}\approx r^{*}$ and both Neel and Brownian relaxation 
contribute and (c) $\bar{r_{c}}>r^{*}$ and Brownian relaxation dominates. The plots exhibit a two-humped 
structure when both relaxation times contribute (case b). The $\beta$ and the $\tau_{c}$ values 
obtained from the Cole-Davidson equations (refer text in section II C) are also indicated in the figures.\\ 

\noindent{\bf Figure 4:} Variation of $\sigma$ as a function of $\beta$ for three different anisotropy
constants on a semi-logarithmic scale. \\

\newpage
\noindent{\bf Figure 5:} The reconstructed particle size distribution along with the particle size 
distribution obtained using TEM for comparisons (refer text in section II D for details).\\

\noindent{\bf Figure 6:} (a) Variation of the mean cluster size $<k(R,t)>$ as a function of time $t$ for 
$\mu$ $=$ 2 and different values of the ratio $R$ specified in the figure. The inset shows the steady-state 
value of the mean cluster size $<k(R)>$ as a function of $R$ on a double logarithmic scale. The best fit line
to data has a slope $0.85\pm 0.02$ respectively. (b) Scaled data of (a) with the scaled axes as indicated in the
figure. The initial growth of clusters obeys a power law characterized by the exponent $z = 0.38\pm 0.02$.
Similar data corresponding to $\mu$ $=$ 1 is also indicated in the inset. For this data, the growth exponent
$z = 0.78\pm 0.02$ as indicated.\\

\noindent{\bf Figure 7:} (a) Steady-state distribution $P(k)$. vs $k$ for $\mu$ $=$ 2 for indicated values 
of ratio $R$. (b) Scaled data of (a) with the scaled axis as indicated in the figure.\\

\noindent{\bf Figure 8:} Data from reference \cite{promis} on average chain length as a function of
time for different volume fractions replotted on scaled axes as shown in the figure. The 
solid line corresponds to a slope of $0.78$ obtained from a numerical solution of the aggregation-fragmentation 
model with $\mu$ $=$ 1 (refer inset of figure 6b and corresponding text).

\newpage

\vspace{2cm}
\begin{center}
{\bf Table I}\\

\vspace{0.5cm}
\begin{tabular}{|c|c|ccc|ccc|}\hline\hline
$r_{c}$ $(nm)$ & $\tau_{N}$ (s) & \multicolumn{3}{c|}{\emph{$\tau_{B}$ (s)}} & \multicolumn{3}{c|}{\emph{$\tau_{e}$ (s)}}\\
         & (s) & $\delta_{1}$=2 $nm$ & $\delta_{2}$=4 $nm$& $\delta_{3}$=6 $nm$& $\tau_{e_{1}}$& $\tau_{e_{2}}$& $\tau_{e_{3}}$\\  \hline
 4  & $10^{-9}$& $10^{-7}$ & $10^{-6}$ & $10^{-6}$ & $10^{-9}$ & $10^{-9}$ & $10^{-9}$  \\ \hline
 6  & $ 10^{-8}$ & $10^{-6}$ & $10^{-6}$ & $10^{-6}$ & $10^{-8}$ & $10^{-8}$ & $10^{-8}$ \\ \hline
8  & $10^{-5}$ & $10^{-6}$ & $10^{-6}$ & $10^{-6}$ & $10^{-6}$ & $10^{-6}$ & $10^{-6}$ \\ \hline
10  & $ 10^{-1}$ & $10^{-6}$ & $10^{-6}$ & $10^{-5}$ & $10^{-6}$& $10^{-6}$ & $ 10^{-5}$ \\ \hline
15  & $10^{20} $ & $10^{-5}$ & $10^{-5}$ & $10^{-5}$ & $10^{-5}$ & $10^{-5}$ & $10^{-5}$  \\ \hline
20  & $10^{61}$ & $10^{-5}$ & $10^{-5}$ & $10^{-5}$ & $10^{-5}$ & $10^{-5}$ & $10^{-5}$ \\ \hline\hline
\end{tabular}

\vspace{2cm}
{\bf Table II}\\

\vspace{0.5cm}
\begin{tabular}{|l|c|c|c|c|c|r|}\hline\hline
MNP Sample & Core & Shell & $\Omega$ & $<k>$ \\  \hline
DDN128 & $ Fe_{3}O_{4}$ & carboxydextran & 0.5364 & 2.3 \\
\hline
FluidMagD5 & $Fe_{3}O_{4}$ & starch & 0.36 & 2.0\\  \hline
MagBSA & $Fe_{3}O_{4}$ & BSA & 0.676 & 2.5\\
\hline
Resovist & $ Fe_{3}O_{4}$ & Carboydextran & 0.51 & 2.4\\
\hline\hline
\end{tabular}
\\

\end{center}

\newpage

\vspace{2cm}
\begin {center}

\includegraphics[width=12.0cm,height=10cm,angle=0]{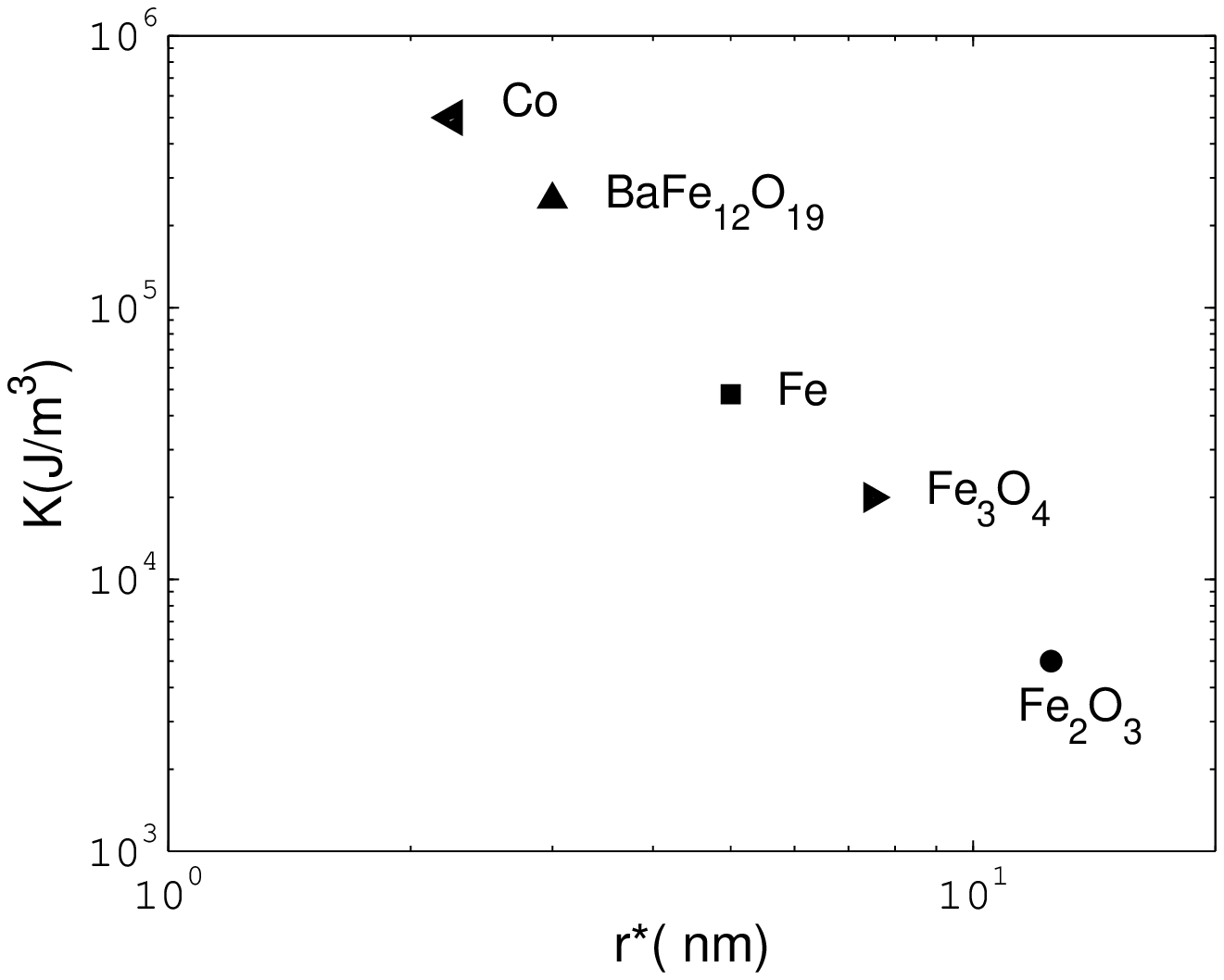}\\
{\bf Figure 1}\\

\vspace{1.0cm}
\includegraphics[width=12.0cm,height=10cm,angle=0]{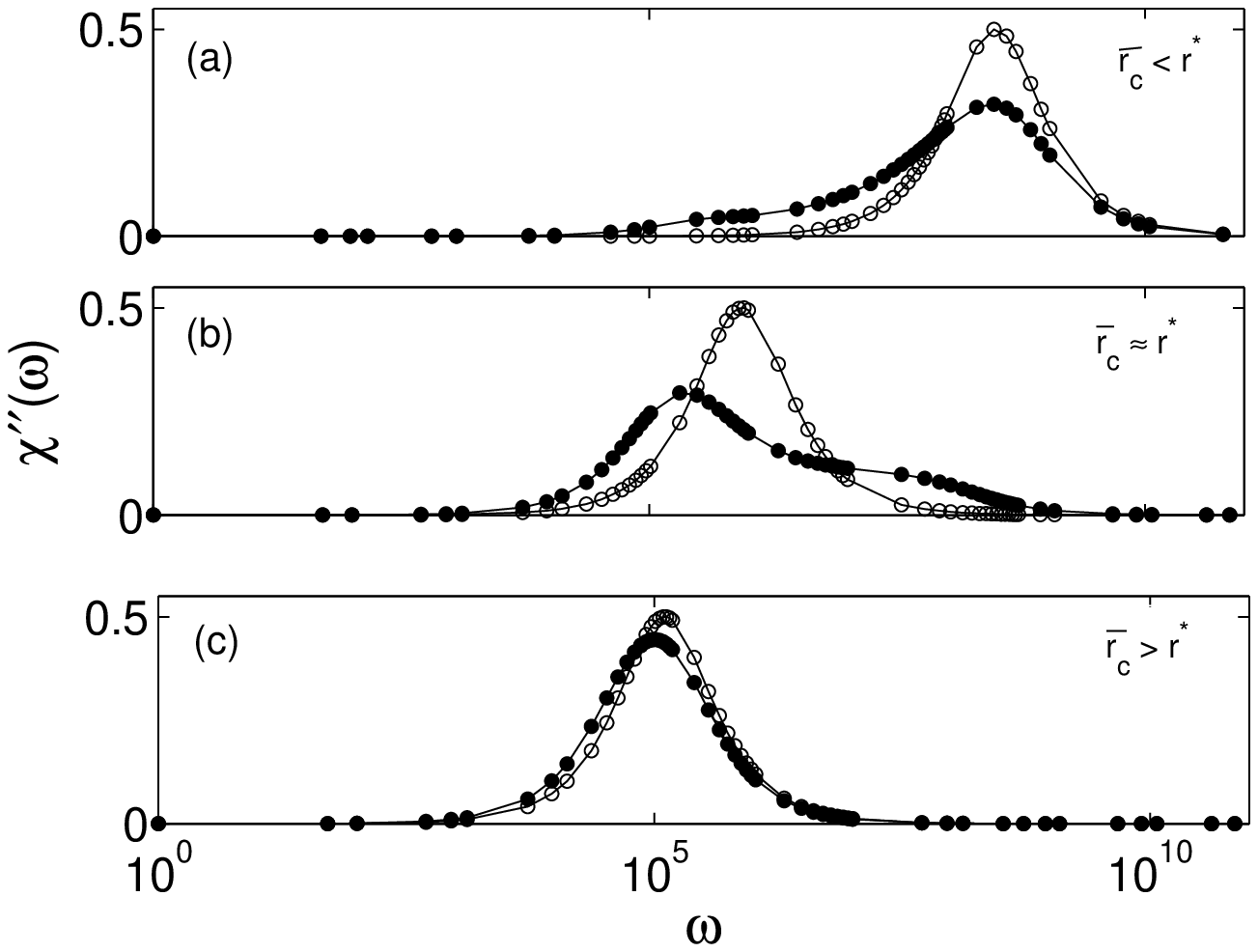}\\
{\bf Figure 2}
\end{center}

\newpage

\vspace{2cm}
\begin {center}

\includegraphics[width=12.0cm,height=10cm,angle=0] {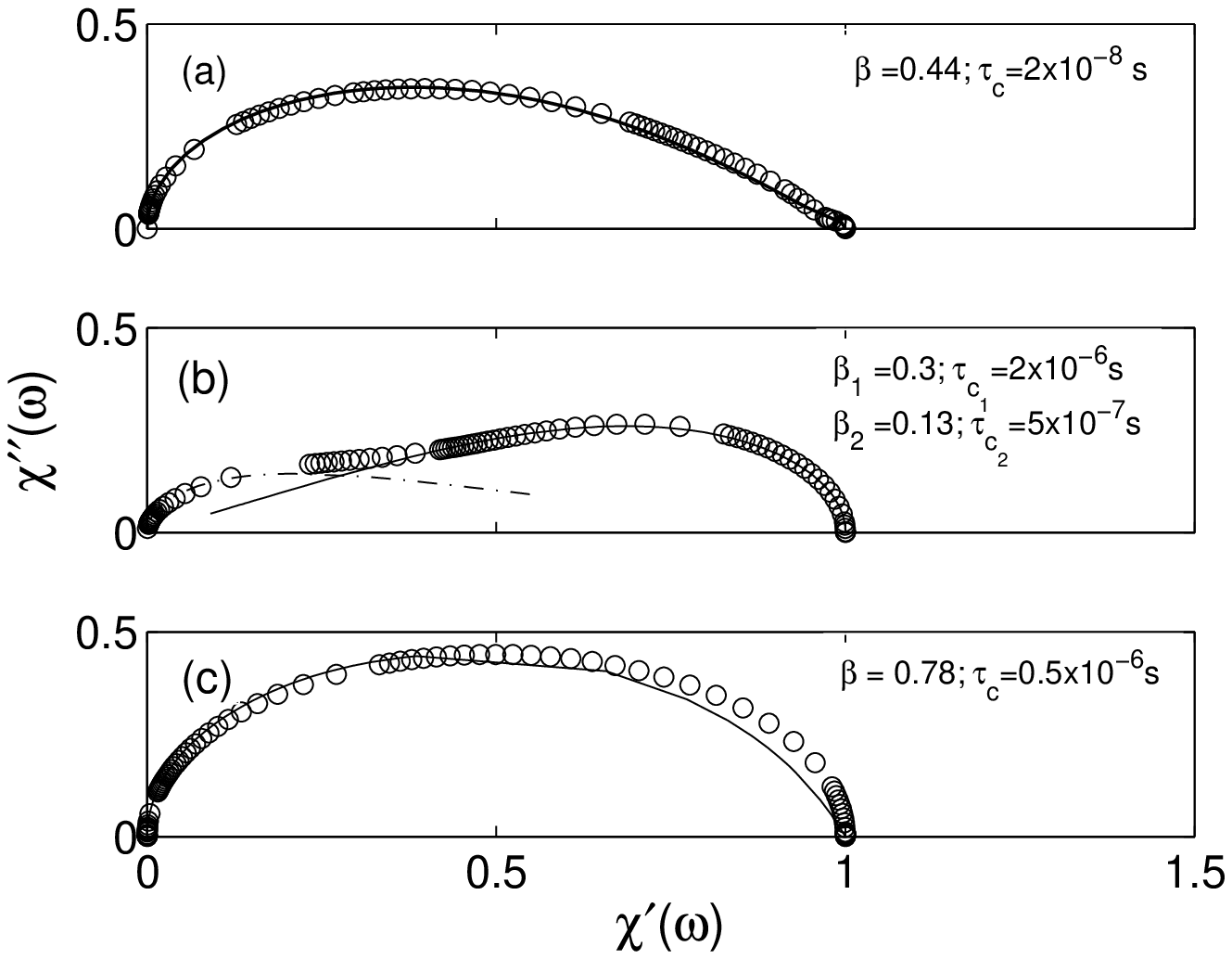}\\
{\bf Figure 3}\\

\vspace{1cm}
\includegraphics[width=12.0cm,height=10cm,angle=0]{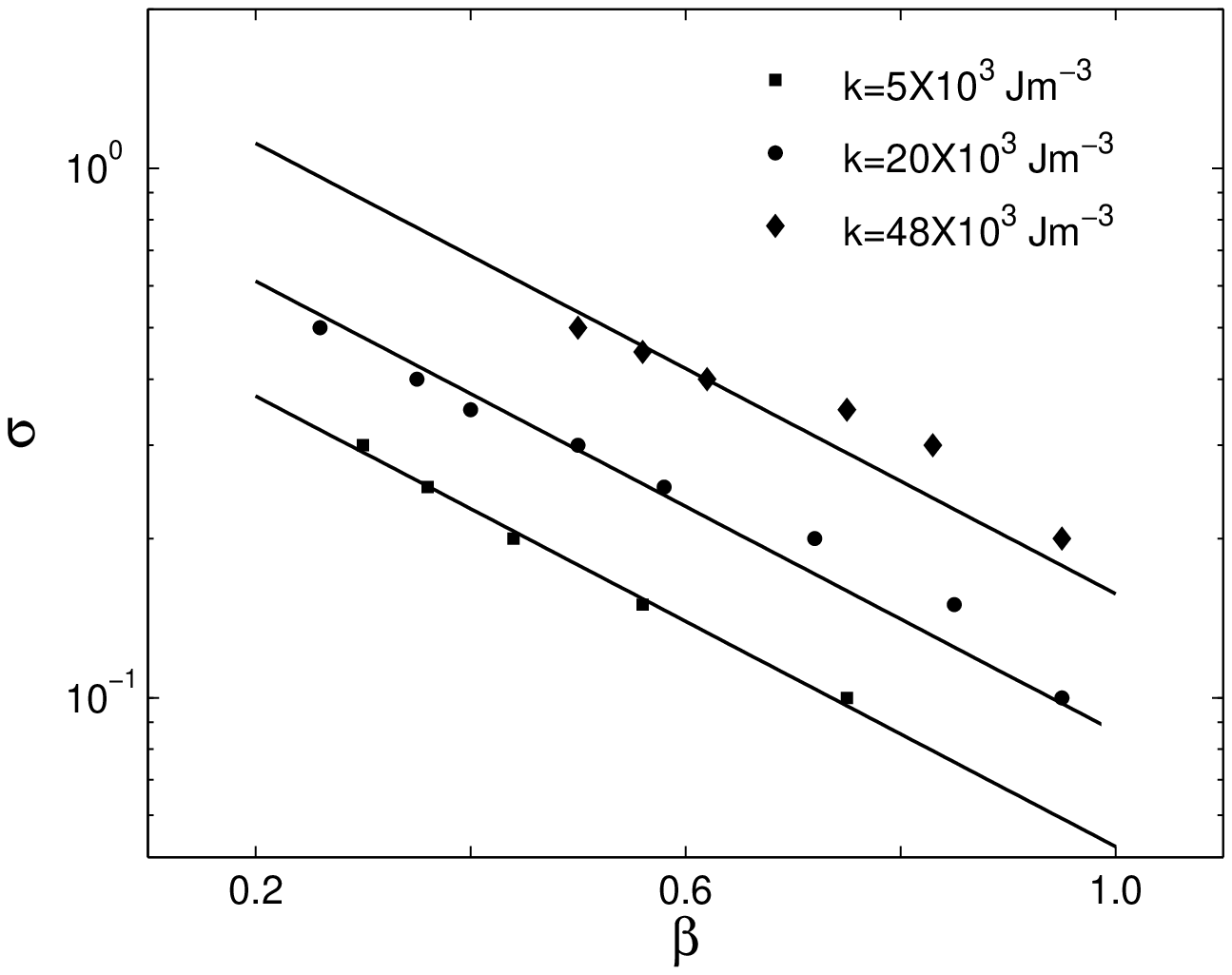}\\
{\bf Figure 4}
\end{center}

\newpage

\vspace{2cm}
\begin {center}

\includegraphics[width=12.0cm,height=10cm,angle=0]{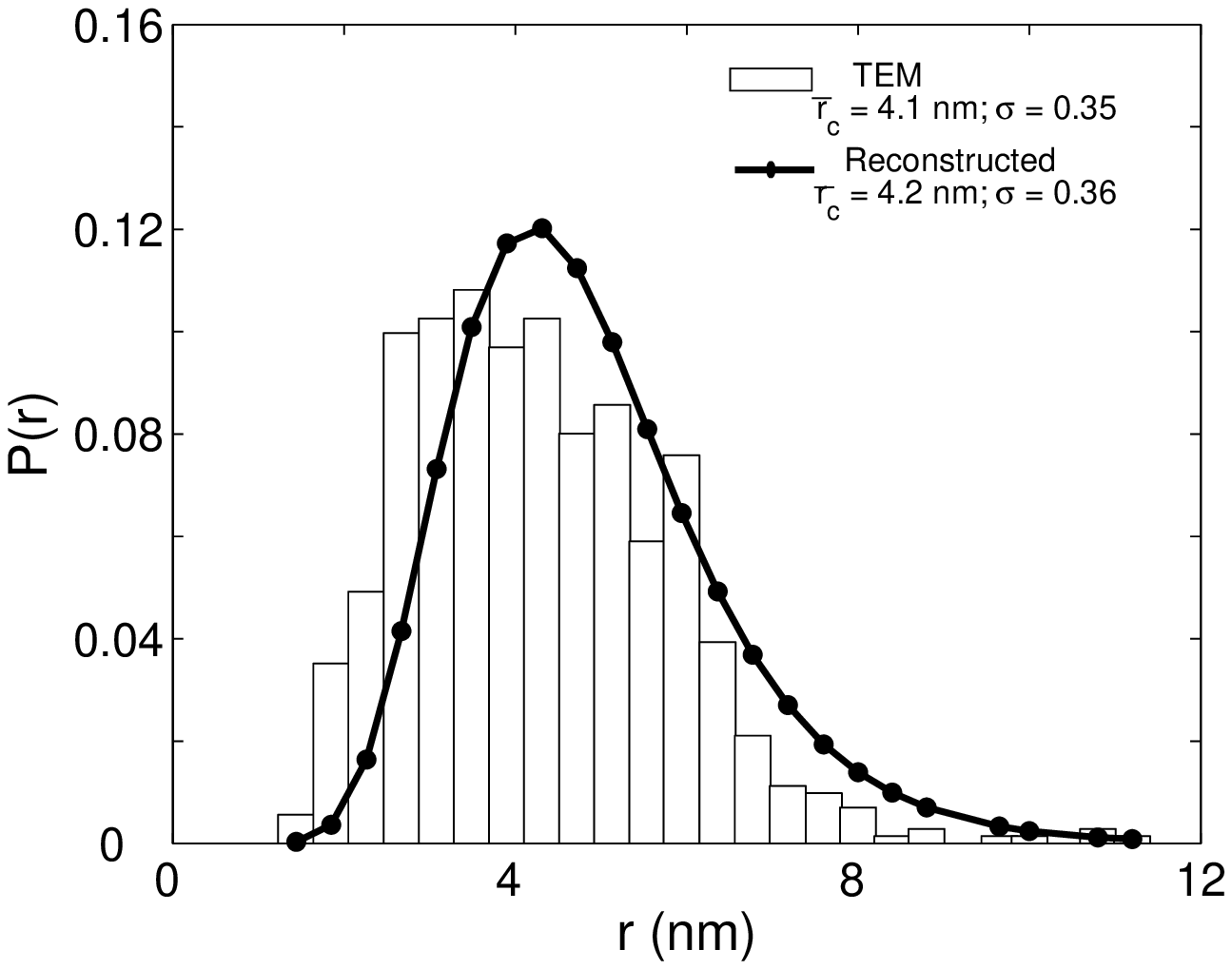}\\
{\bf Figure 5}\\
\vspace{1cm}

\includegraphics[width=12.0cm,height=10cm,angle=0]{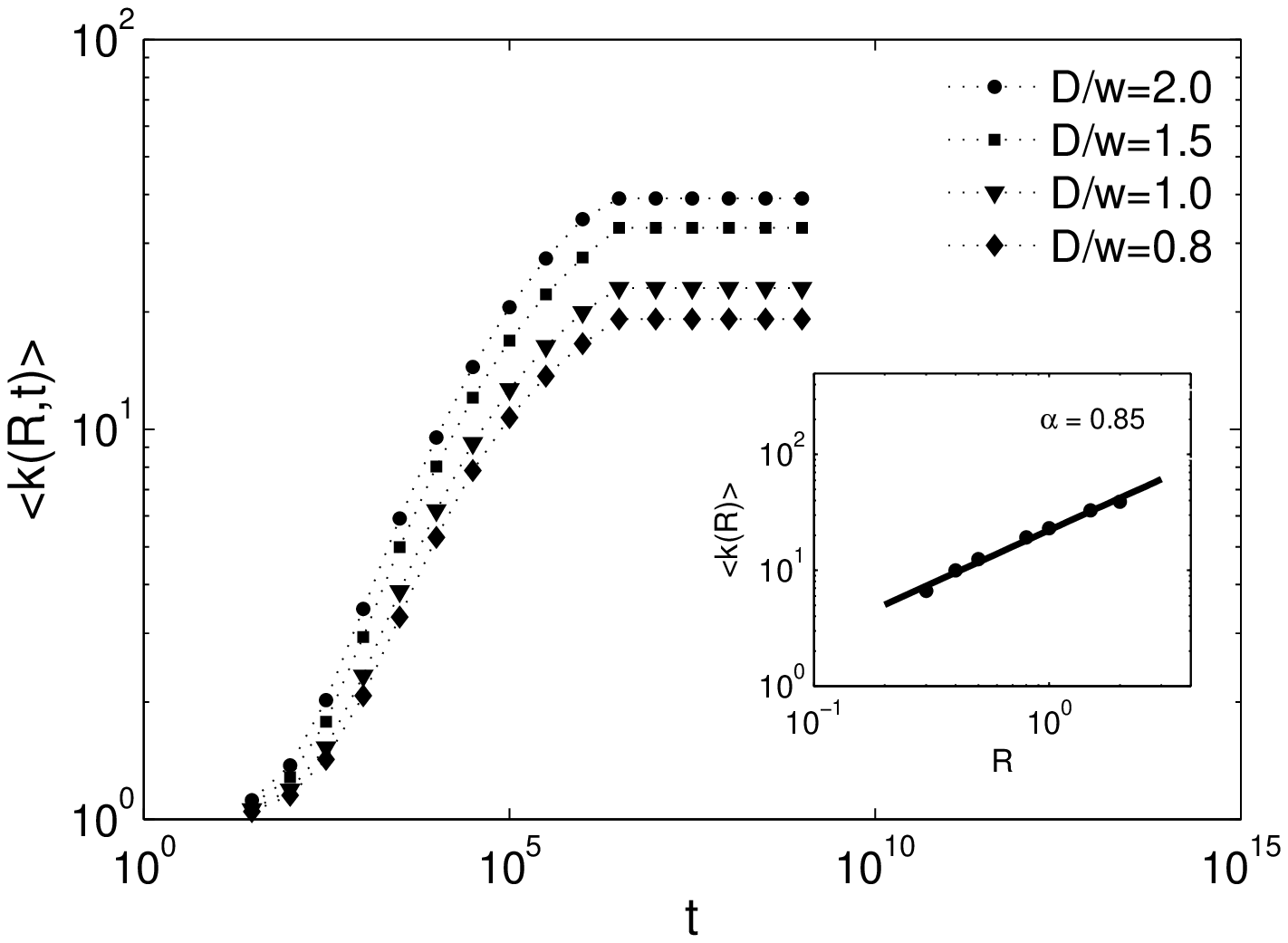}\\
{\bf Figure 6a}\\

\newpage
\vspace{2cm}
\includegraphics[width=12.0cm,height=10cm,angle=0]{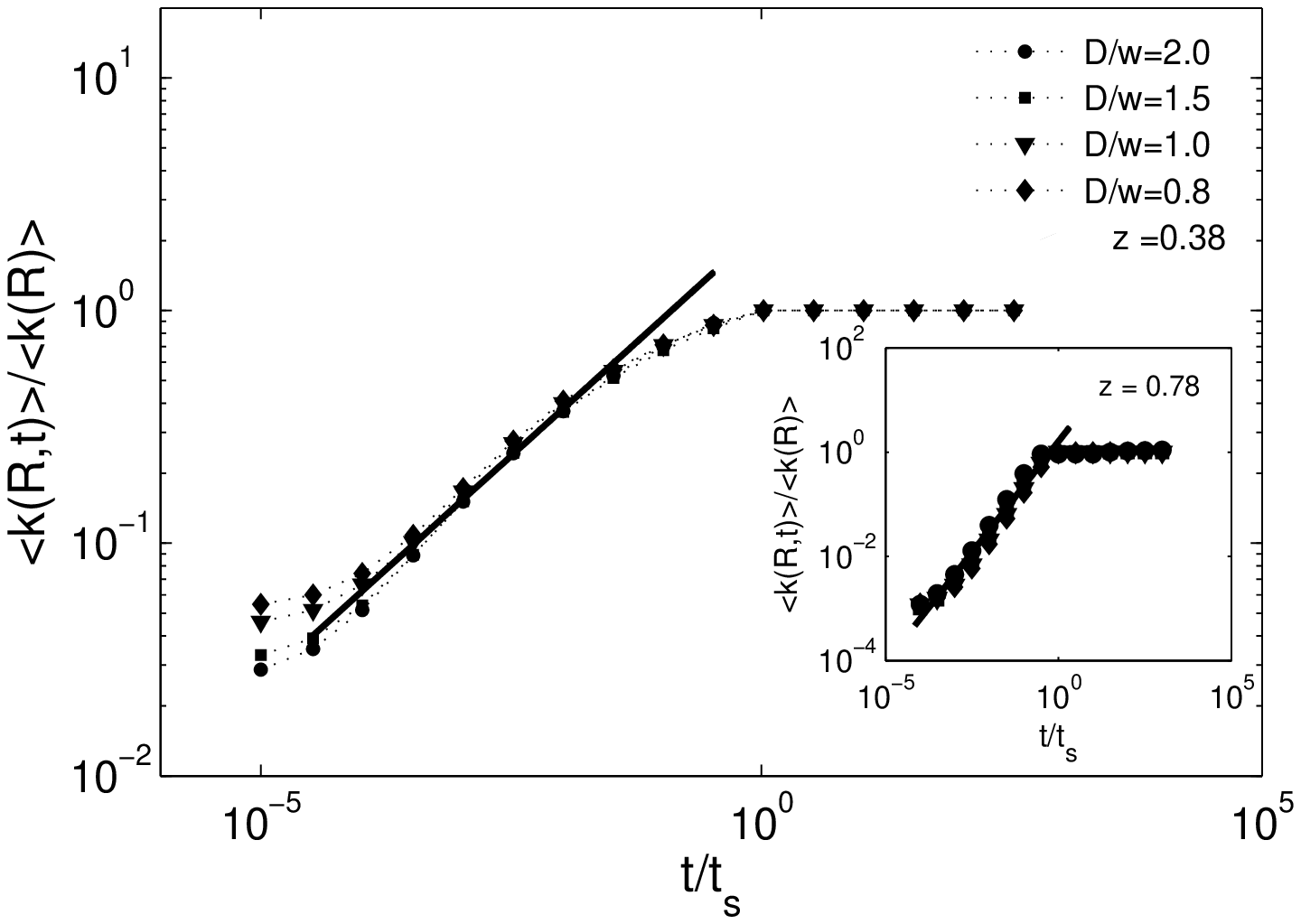}\\
{\bf Figure 6b}

\vspace{1cm}
\includegraphics[width=12.0cm,height=10cm,angle=0]{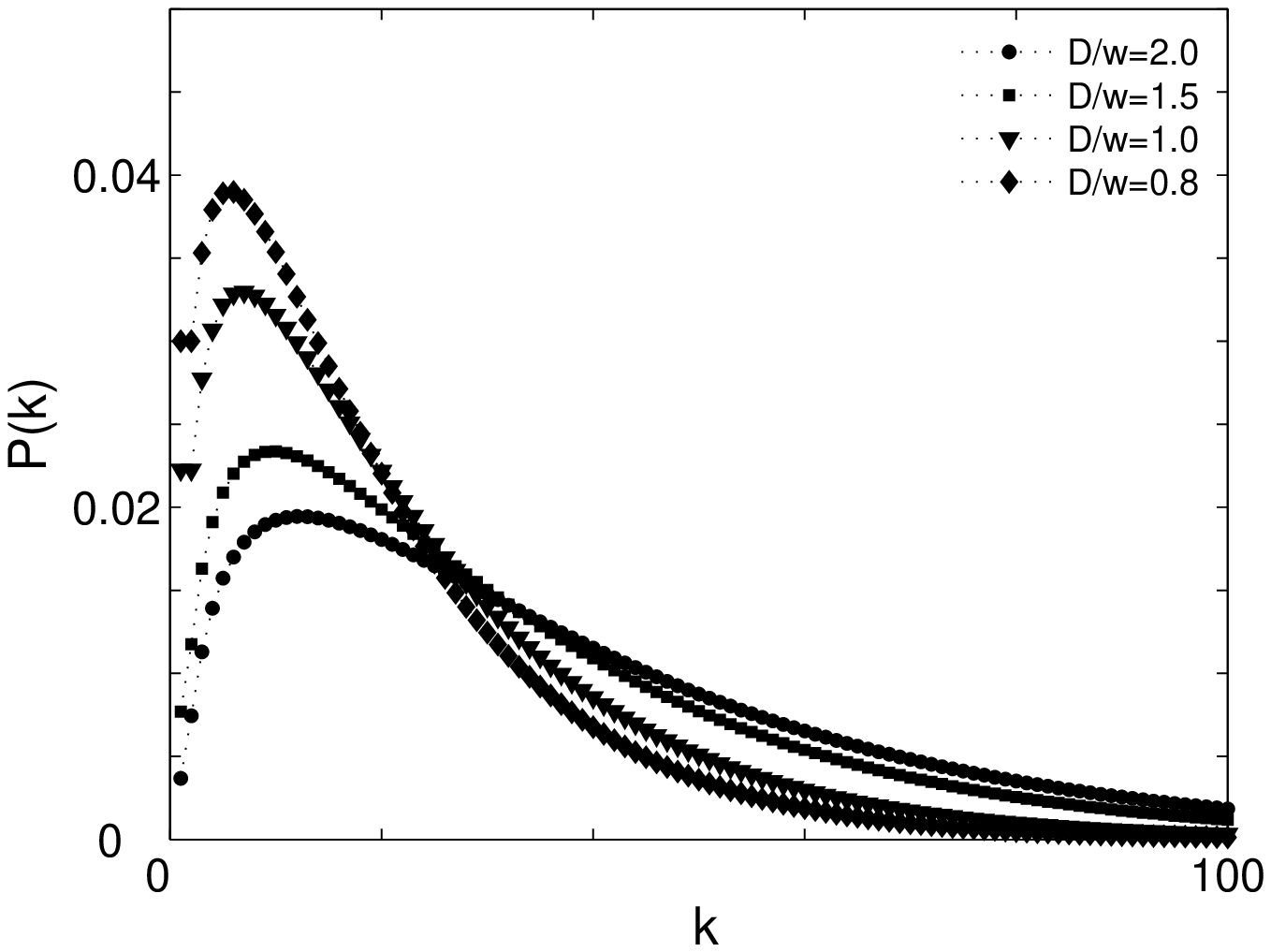}\\
{\bf Figure 7a}\\

\newpage

\vspace{2cm}
\includegraphics[width=12.0cm,height=10cm,angle=0]{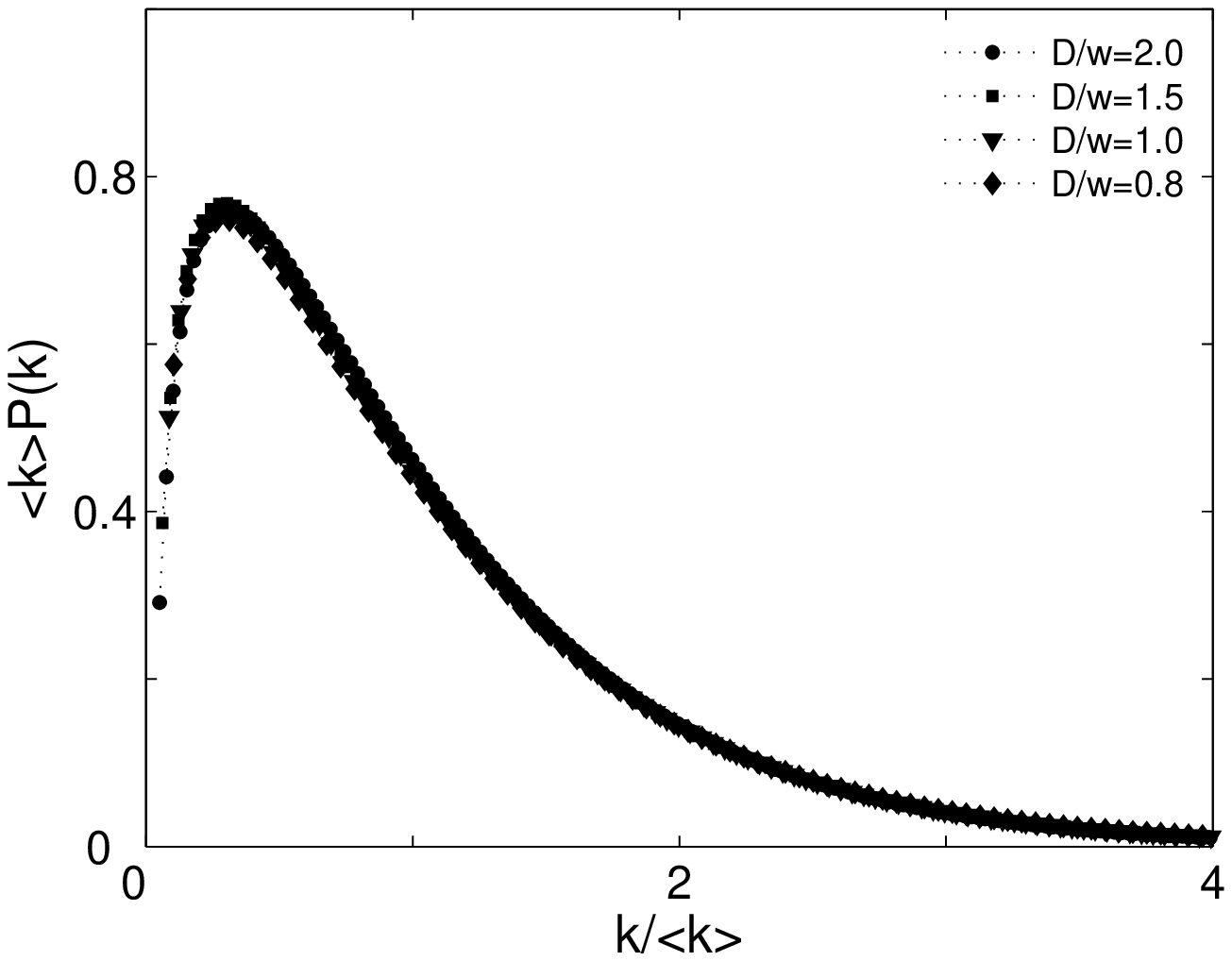}\\
{\bf Figure 7b}

\vspace{1cm}
\includegraphics[width=12.0cm,height=10cm,angle=0]{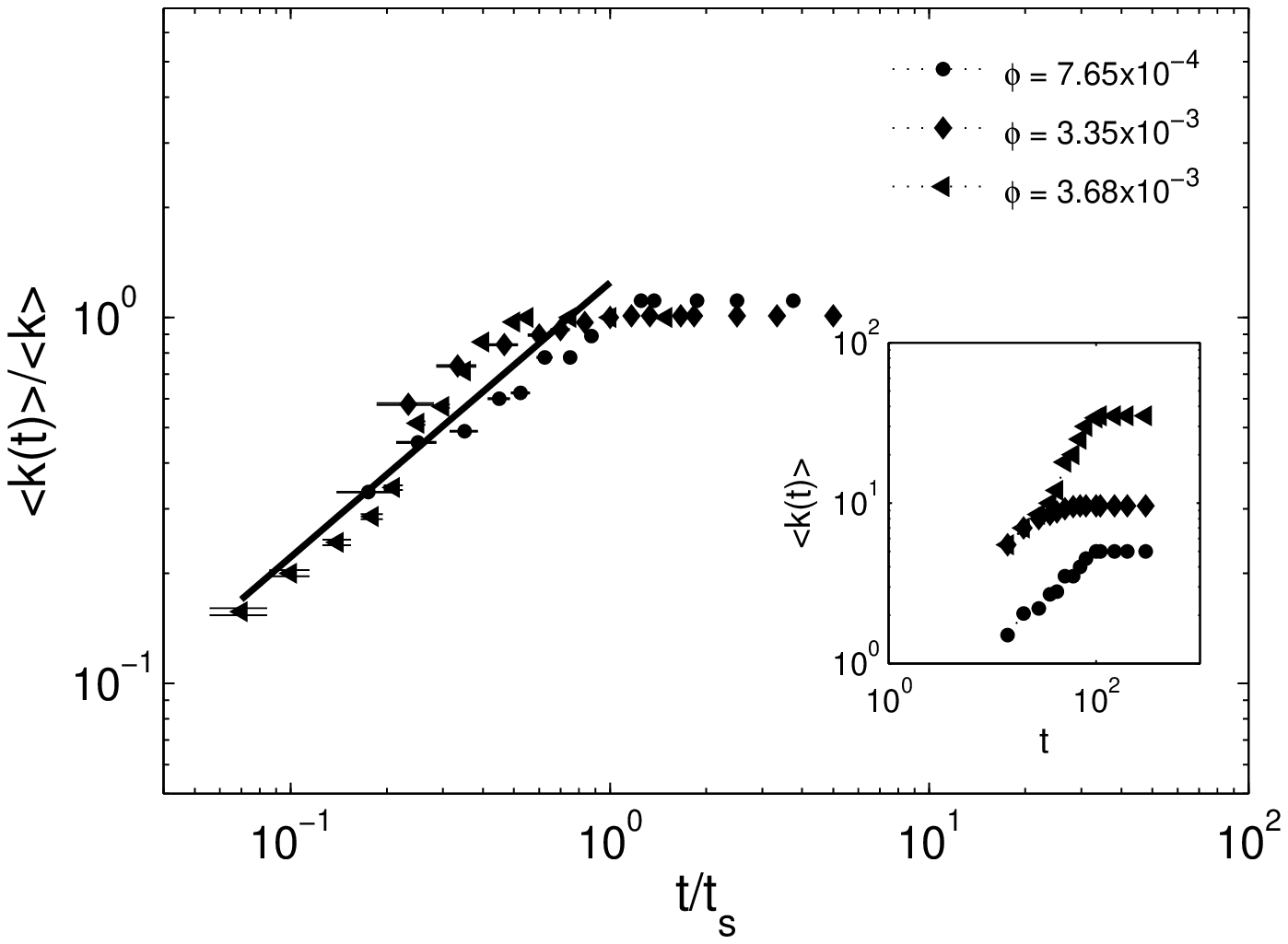}\\
{\bf Figure 8}
\end{center}


\newpage
\begin{thebibliography}{99}
\bibitem{rosen} { Rosensweig R E 1997 {\it Ferrohydrodynamics} (New York: Dover)}.
\bibitem{odenb} { Odenbach S 2002 {\it Magnetoviscous effects in ferrofluids} (Berlin: Springer Verlag)}
\bibitem{coffey} { Coffey W T, Kalmykov Yu P and Waldron J T 1996 {\it The Lagevin Equation} (Singapore: World Scientific)}
\bibitem{conno} { Pankhurst Q A, Connolly J, Jones S K and Dobson J 2003 {\it J.Phys.D: Appl.Phys.} {\bf 36} R167 }
\bibitem{astal} { Astalan A P, Ahrentrop F, Johansson C, Larsson K and Krozer A 2004 
{\it Biosensors and Bioelectronics} {\bf 19} 945}
\bibitem{chung} {Chung S, Hoffmann A and Bader S A 2004 {\it Appl. Phys. Lett.} {\bf 85} 2971}
\bibitem{shlio} { Shliomis M I and Raikher Y 1980 {\it IEEE Mag.} {\bf 16} 237 }
\bibitem{fannin1} { Fannin P C and Charles S W 1989 {\it J. Phys. D: Appl. Phys.} {\bf 22} 187 }
\bibitem{rinal} { Rinaldi C, Chaves A, Elborai S, He X and Zahn M 2005 {\it Current Opinion in Colloid and Interface Sciences} {\bf 10} 141}
\bibitem{cole1} { Cole K S and Cole R H 1941 {\it J. Chem. Phys.} {\bf 9} 341}
\bibitem{cole2} { Cole K S and Cole R H 1942 {\it J. Chem. Phys.} {\bf 10} 98}
\bibitem{zhaom} { Zhao M, Kircher M F, Josephson L and Weissleder R 2002 {\it Bio. Conjug. Chem.} {\bf 13} 840}
\bibitem{promis} { Promislow J H E, Gast A P and Fermigier M 1995 {\it J. Chem. Phys.} {\bf 102} 5492}
\bibitem{martin} { Hagenbuchle M and Liu J 1997 {\it Appl. Optics} {\bf 36} 7664}
\bibitem{kellner} { Kellner R R and Kohler W 2005 {\it J. Appl. Phys.} {\bf 97} 034910}
\bibitem{hasmo} { Hasmonay E and Depeyrot J 2000 {\it J. Appl. Phys.} {\bf 88} 6628}
\bibitem{eber1} { Eberbeck D, Wiekhorst F, Steinhoff U and Trahms L 2006 {\it J. Phys.: Condens. Matter} {\bf 18} S2829}
\bibitem{smolu} { Smoluchowski M V 1916 {\it Z.Phys.} {\bf 17} 585}
\bibitem{chandra} { Chandrasekhar S 1943 {\it Stochastic Problems in Physics and Astronomy, Rev. Mod. Phys.} {\bf 15} 1}
\bibitem{meakin} { Miyazima S, Meakin P and Family F 1987 {\it Phys. Rev.} A {\bf 36} 1421}
\bibitem{kandel} { Kandel D 1997 {\it Phys. Rev. lett.} {\bf 79} 4238}
\bibitem{morim} { Morimoto H and Maekawa T 2000 {\it J. Phys. A: Math. Gen.} {\bf 33} 247}
\bibitem{sdgupta} { Dattagupta S 1987 {\it Relaxation Phenomena in Condensed Matter Physics}  (Orlando: Academic Press)}
\bibitem{brown} { Brown W F 1963 {\it J. Appl. Phys.} {\bf 34} 1319}
\bibitem{chika} { Chikazumi S 1997 {\it Physics of Ferromagnetism} (New York: Oxford University Press )}
\bibitem{fannin2} { Fannin P C, Relihan T and Charles S W 1997 {\it Phys.Rev} B {\bf 55} 14423}
\bibitem{coldavid} { Davidson D W and Cole R H 1951 {\it J. Chem. Phys.} {\bf 19} 1484}
\bibitem{ernst} { Ernst M H and Dongen P G J 1987 {\it Phys. Rev.} A {\bf 36} 435}
\bibitem{krapi} { Krapivsky P L and Redner S 1996 {\it Phys. Rev.} E {\bf 54} 3553}
\bibitem{chavez} { Chavez F, Moreau M and Vicente L 1997 {\it J. Phys. A.: Math. Gen.} {\bf 30} 6615}
\bibitem{kalachev} { Kalachev L, Morimoto H and Maekawa T 2001 {\it Int. J. Mod. Phys. B} {\bf 15} 774}
\bibitem{eber2} { Eberbeck D, Bergemann C, Hartwig S, Steinhoff U and Trahms L 2005 
{\it J. Magn. Magn. Mater.} {\bf 289} 435}

\end{thebibliography}
\end{document}